\documentclass{iau-JDSS}
\usepackage{graphicx}

\title[Dwarf Galaxiesand Globular Clusters] 
{Dwarf Galaxies and Globular Clusters}

\author[M. Bellazzini]{Michele Bellazzini$^1$}%

\affiliation{$^1$INAF - Osservatorio Astronomico di Bologna, via Ranzani 1, 40127, Bologna, Italy
\break email: michele.bellazzini@oabo.inaf.it\\}

\pubyear{2013}
\volume{Volume xx}  
\pagerange{yyy--zzz}
\date{?? and in revised form ??}
\jname{Highlights of Astronomy, Volume xx}
\editors{Gianpaolo Piotto and Enrico Vesperini, eds.}
\begin{document}

\maketitle

\begin{abstract}
I briefly explore some relevant connections and differences between the evolutionary paths of dwarf galaxies and globular clusters. 

\end{abstract}

Two decades of deep photometric and spectroscopic studies of nearby dwarf galaxies has revealed that significant differences in the Star Formation (SF) and chemical evolution histories can be observed also among gas poor and predominantly old dwarf spheroidals (see \cite[Tolstoy et al. 2009]{eline}).
Still, a recent systematic study of 60 nearby dwarf galaxies showed that the average dwarf galaxy formed more than half of its stars by z=2, i.e. in the first 2-3 billion years from the big bang, irrespectively of its morphological type (\cite[Weisz et. al 2010]{weisz}). It is very interesting to note that {\em (a)} this was also the epoch when the large majority of the Galactic globular clusters (GC) were formed (\cite[Dotter et al. 2010]{dotter}), and {\em (b)} during which Milky Way (MW) sized galaxies are predicted to assemble most of their mass by hierarchical merging, according to the more recent $\Lambda$-CDM simulations (\cite[Abadi et al. 2003]{abadi}, \cite[Fakhouri et al. 2010]{fak}). The fact that, in average, half of the stellar mass of dwarfs was produced in the earliest 2-3~Gyr and half in the following 10~Gyr, indicates that SF proceeded at a very different pace in two phases: it was very fast in the first phase, with chemical enrichment dominated by SNII and  production of metal-poor stars with enhanced abundance of $\alpha$ elements, and slower (and more discontinuous) in the second phase, with self-enrichment dominated by SNI and production of more metal-rich stars and [$\alpha$/Fe] declining with metallicity. It was in this second long-lasting phase that most of the observed dwarf-to-dwarf differences in the SF history took place. 

Hence, dwarf galaxies - in their earliest times - appear as the ideal environment for the birth of GCs, also providing the gas pre-enriched in metals and $\alpha$-enhanced that was required to produce the GCs that we observe today (there is no known GC with [Fe/H]$<-2.5$, hence none of them was formed from pristine gas). Now we have direct proofs that at least a significant fraction of GCs in MW-sized galaxies comes from disrupted/disrupting satellites (\cite[Bellazzini et al. 2003]{mic}, \cite[Law \& Majewski 2010]{lm10}, \cite[Mackey et al. 2010]{mackey}).

\begin{figure}
\center
 \includegraphics[width=9cm]{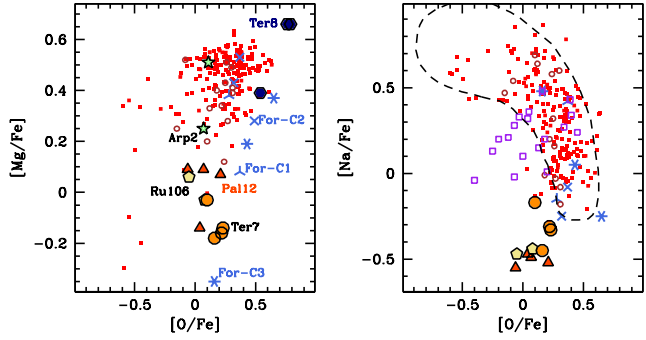}
  \caption{Distributions of Mg vs. O and Na vs. O abundances for stars in different star clusters.
  Small red dots are stars from 15 MW GCs (\cite[Carretta et al. 2009]{c09}); large symbols (labelled) are stars from GCs in dSph satellites of the MW (Fornax and Sagittarius) plus Ru106 (\cite[Letarte et al. 2006]{letarte}, \cite[Brown et al. 1997]{brown}, \cite[Cohen 2004]{judy}, \cite[Mottini et al. 2008]{mottini}, \cite[Sbordone et al. 2005]{sbor}). Small open circles are stars from LMC GCs (\cite[Mucciarelli et al. 2010]{muccia}). Open squares are {\em mean} abundances of a set of open clusters from \cite{desilva}; the dashed contour enclose the distribution of stars in the GC M54, in the central nucleus of Sgr (\cite[Bellazzini et al. 2008]{mic54}, \cite[Carretta et al. 2010b]{c10b}).}
  \label{proc}
\end{figure}


The recent discovery that also GCs were able to produce more than one generation of stars, with 
chemical self-enrichment, may be seen as an unexpected similarity between GCs and dwarf galaxies. In fact this marks a deep difference in the evolutionary path of the two classes of stellar systems:{\em (a)} the chemical evolution of dwarf galaxies was driven by supernovae, leading to correlated enrichment in iron, $\alpha$ elements, etc., while {\em (b)} the chemical evolution of the overwhelming majority of GCs was driven by low-energy polluters, leading to significant (and correlated) spread in the abundance of a few light elements (He, C, N, Na, O, Mg, Al) and {\em no enrichment in iron} (\cite[Gratton et al. 2012]{grat}). The specific abundance pattern observed in GCs has been suggested as a {\em defining property} for these systems (\cite[Carretta et al. 2010b]{c10b}). This is a powerful approach, but the comparisons shown in Fig.~\ref{proc} suggest that it may not be sufficient to embrace the whole diversity of observed cases. Stars from clusters associated with the disrupting Sgr dSph (as well as Ru~106, likely coming from a completely disrupted satellite) have Na, O, and Mg abundance very different from any other star from Galactic GCs and open clusters (see also K. Venn, these proceedings). The chemical initial conditions set by the evolutionary path of the host dwarf galaxy are likely another important factor in determining the subsequent chemical evolution of GCs.

\begin{acknowledgments}
I acknowledge the financial support of INAF through the PRIN Grant assigned to the project {\em Formation and evolution of massive star clusters}, P.I.: R. Gratton.
\end{acknowledgments}

\end{document}